\documentclass[preprint,preprintnumbers,amsmath,amssymb]{revtex4}
\usepackage{amsfonts}
\usepackage{amssymb}
\usepackage{latexsym}
\usepackage{graphicx}% Include figure files
\usepackage{psfig}
\newcommand{\al}{\alpha}

\newcommand{\be}{\beta}

\newcommand{\si}{\sigma}
\newcommand{\Si}{\Sigma}

\newcommand{\De}{\Delta}

\newcommand{\rar}{\rightarrow}

\begin{document}

\preprint{M\'exico ICN-UNAM 15/06, \
                           October 2006 }

\title{The $He_2^{2+}$ molecular ion can exist in a magnetic field}

\author{A.~V.~Turbiner}
\email{turbiner@nucleares.unam.mx}
\author{N.~L.~Guevara}
\email{nicolais@nucleares.unam.mx}
\affiliation{Instituto de Ciencias Nucleares, Universidad Nacional
Aut\'onoma de M\'exico, Apartado Postal 70-543, 04510 M\'exico,
D.F., Mexico}

%\date{}

\begin{abstract}
 A detailed study of the ground state of the Coulomb system
 $(\al \al e e)$ which corresponds to the $He_2^{2+}$ molecular ion in a magnetic field
 $B=0-4.414 \times 10^{13}\,$\,G in parallel configuration (infinitely massive
 $\al-$particles are situated along a magnetic field line) is presented.
 The variational method is employed using a trial
 function which includes electronic correlation in the form
 $\exp{(\gamma r_{12})}$ where $\gamma$ is a variational parameter.
 It is shown that the quantum numbers of the lowest total energy
 state depend on the magnetic field strength. It evolves from the spin-singlet
 ${}^1\Si_g$ metastable state at $0 \leq B \lesssim 0.85$\,a.u. to a repulsive
 spin-triplet ${}^3\Si_u$ state for $0.85\,\mbox{a.u.} \lesssim \ B \lesssim 1100$\,a.u.
 and, finally, to a strongly bound spin-triplet ${}^3\Pi_u$ state at stronger fields
 $B \gtrsim 1100$\,a.u.
\end{abstract}

%\pacs{31.15.Pf,31.10.+z,32.60.+i,97.10.Ld}

\maketitle

\section{\protect\bigskip introduction}

It is well-known that very strong magnetic fields may appear on the
surface of white dwarfs, $B\approx 10^{6}-10^{9}$\,G and neutron
stars, $B\approx 10^{11}-10^{13}$\,G (for latter case they can be
even stronger). It seems natural to expect that in a vicinity of the
surface atomic and molecular systems can occur. For many years this
assumption motivates a development of atomic-molecular physics in a
strong magnetic field since it might lead to understanding of the
spectra of these compact stars.

So far, a major attention was paid to a study of one-electron atomic
and molecular systems (for s review see e.g. \cite{PR}). During the
last years it was predicted the existence of many exotic, strongly
bound one-electron molecular system. In turn, two-electron systems
had explored very little being mostly focused to a study of
atomic-type systems $H^-, He$ with the only exception of  the $H_2$
molecule (see \cite{turbinerH2}, \cite {schmelcherH2} and references
therein). Just recently, a first detailed study of the molecular ion
$H_3^+$ was carried out \cite{Turbiner:2006}. It manifested quite
surprising evolution of the ground state as a magnetic field
increases. A goal of the present paper is to make a first study of
the Coulomb system $(\al \al e e)$ in a magnetic field and establish
the existence of the molecular ion $He_2^{2+}$ and its excited
states. It is worth mentioning that the molecular ion $He_2^{2+}$
exists in metastable state in field-free case (see e.g.
\cite{pauling}) being characterized by a well-pronounced potential
well at a finite internuclear distance.

Atomic units are used throughout ($\hbar$=$m_e$=$e$=1), although
energies are expressed in Rydbergs (Ry). The magnetic field $B$ is
given in a.u. with $B_0= 2.35 \times 10^9\,G$.

\section{Generalities}

Let us consider the Coulomb system $(\al \al e e)$ placed  in a
uniform constant magnetic field. We assume that the $\al-$particles
are infinitely massive (Born-Oppenheimer approximation of zero
order). They are situated along the magnetic field line (parallel
configuration). The Hamiltonian which describes this system when the
magnetic field is oriented along the $z$ direction, ${\bf
B}=(0,0,B)$ is written as follows
\begin{equation}
 {\cal H}\ =\
 \sum_{\ell=1}^2 \left( {\hat {\mathbf p}_{\ell}+{\cal A}_{\ell}} \right)^2\
 -\ \sum_{\buildrel{{\ell}=1,2}\over{\kappa =A,B}} \frac{4}{r_{{\ell},\kappa}}
 \ +\ \frac{2}{r_{12}}\ +\ \frac{8}{R}\ +\ 2{\bf B} \cdot {\bf S}  ,
\end{equation}
where ${\hat {\mathbf p}_{\ell}}=-i \nabla_{\ell}$ is the 3-vector
of the momentum of the ${\ell}$th-electron, the index $\kappa$ runs
over the $\al-$particles $A$ and $B$, $r_{12}$ is the interelectron
distance and $\bf{S}=\hat s_{1}+\hat s_{2}$ is the operator of the
total spin. ${\cal A}_{\ell}$ is a vector potential which
corresponds to the constant uniform magnetic field $\bf B$ chosen in
the symmetric gauge,
\begin{equation}
   {\cal A}_{\ell}= \frac{1}{2}({\bf{B}} \times \ {\bf{r}}_{\ell})
   = \frac{B}{2} (-y_{\ell},\ x_{\ell},\ 0)\ .
\end{equation}
%%%%%%%%%%%%%%%%%%%%%%%%%%%%%%%%%%%%%%%%%%%%%%%%%%%%%%%%%%%%%%%%%%
After substitution of (2) to (1) we arrive at the Hamiltonian in the
form
\begin{equation}
 %\hspace{-10pt}
  {\cal H}\ =\ \sum_{{\ell}=1}^2 \left(- {\mathbf\nabla}^2_ {\ell}
  \ +\ \frac{B^2}{4} \rho_{\ell}^2 \right)   -\sum_{{\ell},\kappa}
  \frac{4}{r_{{\ell}\kappa}}  + \frac{2}{r_{12}} + \frac{8}{R} + B
  (\hat L_z +2\hat S_z ) \ ,
\end{equation}
where  $\hat L_z=\hat l_{z_1}+\hat l_{z_2}$ and $\hat S_z=\hat
s_{z_1}+\hat s_{z_2}$ are the z-components of the total angular
momentum and total spin, respectively. The variable
$\rho_{\ell}=\sqrt{x_{\ell}^2+y_{\ell}^2}$ is the distance from
$\ell$th electron to $z-$axis.

The problem under study is characterized by three integrals of
motion: (i) the operator of the $z$-component of the total angular
momentum (projection of the angular momentum on the magnetic field
direction) giving rise to the magnetic quantum number $m$, (ii) the
spatial parity operator $P({\vec r_1} \rar -{\vec r_1},{\vec r_2}
\rar -{\vec r_2})$ which has eigenvalues $p=\pm 1$(gerade/ungerade)
(iii) the operator of the $z$-component of the total spin
(projection of the total spin on the magnetic field direction)
giving rise to the spin quantum number $m_s$. Hence, to any
eigenstate three explicit quantum numbers can be assigned: the
magnetic quantum number $m$, the parity $p$ and the spin quantum
number $m_s$. Therefore, the space of eigenstates is split into
subspaces (sectors) with each of them being characterized by
definite values of $m$, $p$ and $m_s$.  It is worth noting the
Hamiltonian $\cal H$ is also invariant with respect to $z_1 \to
-z_1$ and $z_2 \to -z_2$ (z-parity operator $P_z$) with quantum
numbers $\sigma_N=\pm 1$ for positive/negative $z$-parity (this
symmetry accounts for the interchange of the nuclei A and B).

Thus, to classify eigenstates we follow the convention widely
accepted in molecular physics using the quantum numbers $m$, $S$ and
$p$. In particular, the notation is $^{2S+1} M_{p}$, where $2S+1$ is
the spin multiplicity and corresponds to $1$ for singlet ($S=0$) and
$3$ for triplet ($S=1$), for the label $M$  Greek letters $\Si, \Pi,
 \De$ are used that correspond to the states with $|m|=0, 1, 2,...$,
respectively, and the subscript $p$ (the spatial parity quantum
number) will take gerade/ungerade($g/u$) labels describing  positive
$p=+1$ and negative $p=-1$ parity states.  However, there exists a relation
between the quantum numbers corresponding to the $z$-parity
(interchange of nuclei A and B) and the spatial parity:
\[
 p=(-1)^m\ \si_N\ .
\]
We restrict our consideration to the states with $m=0, -1, -2$
because the ground state of a sector with $m>0$ always has larger
total energy than those with $m \leq 0$.

As a method to explore the problem the variational procedure is
used. The recipe of choice of trial functions is based on arguments
of physical relevance (see e.g. \cite{turbinervar}). Eventually, a
trial function for the state with magnetic quantum number $m$ is
chosen in a form
\[
 \psi^{(trial)} = (1+\si_e P_{12})(1+\si_N P_{AB})
\]
\begin{equation}
\label{ansatz}
 \rho_1^{\mid m \mid}e^{im\phi_1} e^{
 -\al_1 r_{1A}-\al_2 r_{1B} -\al_3 r_{2A} -\al_4 r_{2B}
 + \gamma r_{12}
 - B  \be_{1}\rho_1^2/4 - B  \be_{2}\rho_2^2/4} \ ,
 \end{equation}
where $\si_e=1,-1$ stands for spin singlet ($S=0$) and triplet
states $(S=1)$, respectively, when $\si_N=1,-1$ stands for nuclear
gerade and ungerade states, respectively. The $P_{12}$ is the
operator which interchanges electrons (1 $\leftrightarrow$ 2) and
$P_{AB}$ is the operator which interchanges the two nuclei $A$ and
$B$. The parameters $\al_{1-4}$, $\be_{1-2}$ and $ \gamma$ are
variational parameters. If the internuclear distance $R$ is taken
into account as a variational parameter the trial function
(\ref{ansatz}) depends on eight parameters.

Calculations were performed using the minimization package MINUIT
from CERN-LIB. Multidimensional integration was carried out using a
dynamical partitioning procedure: the domain of integration was
divided into subdomains following an integrand profile and then each
subdomain was integrated separately (for details see e.g.
\cite{Turbiner:2006}). Numerical integration was done with a
relative accuracy of $\sim 10^{-6} - 10^{-7}$ by use of the adaptive
D01FCF routine from NAG-LIB. Computations were performed on a dual
DELL PC with two Xeon processors of 2.8\,GHz each (ICN), 54-node
FENOMEC and 32-node TOCHITL clusters (UNAM) and DUKE dual DELL PC
with two Xeon processors of 3.06\,GHz each (CINVESTAV).

\newpage

\section{Results}
It is known that in field-free case the $He_2^{2+}$ molecular ion
exists in the state ${}^1\Si_g$ ($S = 0$, $\si_N =1 $ and $m=0$) as
a metastable state - it decays to $He^+ + He^+$ (see e.g.
\cite{kolos}). A natural goal is to check the existence of bound
state of the system $(\al \al e e)$ in parallel configuration in
presence of the magnetic field ranging $B=0-4.414 \times
10^{13}\,$G. To carry out such a study the trial function
(\ref{ansatz}) with $\si_e =1$, $\si_N =1$ and $m=0$ is used, which
depends on 8 variational parameters. The calculations indicate
clearly the existence of a minimum in the total energy curve
$E_T(R)$ for all magnetic fields $B \geq 0$. The concrete results
for the $^1\Si_g$ state are presented in Table~\ref{table1}. This
state is a short-lived state for weak magnetic fields where its
total energy curve lies above the total energy in the dissociation
channel to two separate helium ions $E_T(He^+(1s_0)+He^+(1s_0))$,
for which the spins of electrons in both $He^+$ are antiparallel to
the magnetic field direction. In general, as the magnetic field
strength grows, the total energy increases, the system becomes more
bound - the double ionization energy $E_I=2B-E_T$ increases and more
compact (the internuclear equilibrium distance decreases).

%It is worth mentioning that for the magnetic field $B=100$\,a.u.
%(see Table~\ref{table1}) a comparison of the total energy of
%$He_2^{2+}$ in the ${}^1\Si_g$ state and the total energy of the
%decay channel $He_2^{2+}  \to He^+(1s_0) + He^+(1s_0)$ (since
%$E_T(He_2^{2+} )< 2 E_T(He^+(1s_0)))$ indicates that the $He_2^{2+}$
%molecular ion in the ${}^1\Si_g$ state becomes stable with respect
%to this channel of decay. However, this observation does not make
%much sense - the $He_2^{2+}$ ion does not exist for this magnetic
%field(!), see below.

\begin{table}
  \centering
\caption{\label{table1} Total $E_T$, double ionization $E_I$
  energies (in Ry) and equilibrium distance (in a.u.)
  of the $He_2^{2+}$ ion in the state  $^1\Si_g$. Magnetic fields
  for which $He_2^{2+}$ does not exist (see text) are marked by
  ${}^{(*)}$.
  Total energy $E_T(He^+(1s_0) + He^+(1s_0))$ (in Ry) for zeroing
  total electron spin projection from
  \cite{Turbiner:2006} is shown for a comparison. }
\begin{tabular}{|c|c|c|c||c|}
\hline\hline
B(a.u.)     & $E_T$      &$E_I $&$R_{eq}$ & $2E_T(He^+(1s_0))$  \\
\tableline
0          & -7.353     &       & 1.334   &  -8.0 \\
           & -7.363\footnote{See Ref.\cite{kolos}} & & 1.33$^a$ & \\
$1^{(*)}$          & -7.137     &  9.137  & 1.278  & -7.764     \\
%          &            &&        &               \\
$10^{(*)}$         &  2.628     &  17.372 &  0.841 & 2.440  \\
%          &            &&        &               \\
$100^{(*)}$        & 159.08     &  40.92  &  0.393 & 161.78  \\
%          &            &&        &               \\
$1000^{(*)}$       & 1905.83    &  94.17  &  0.202 & 1919.20 \\
%          &            &&        &               \\
10000      & 19799.3    & 200.7   &  0.094 & 19843.2 \\
%          &            &&        &    \\
\tableline\hline
\end{tabular}
\end{table}

As a next step a detailed study for the  $^3\Si_u$($S = 1$, $\si_N
=-1,m=0 $ ) state of the $He_2^{2+}$ molecular ion in parallel
configuration in the domain of magnetic fields $0~\le~B~ \le~4.414
\times 10^{13}\,$G is carried out.  For this state the variational
trial function (\ref{ansatz}) is characterized by $\si_e =-1$,
$\si_N =-1$ and $m=0$, it depends on 8 variational parameters.

%%%%%%%%%%%%%%  FIGURE:1  %%%%
%%%%%%%%%%%%%%
\begin{figure}[tb]
\begin{center}
   \includegraphics*[width=3.in, height=4.in, angle=-90]{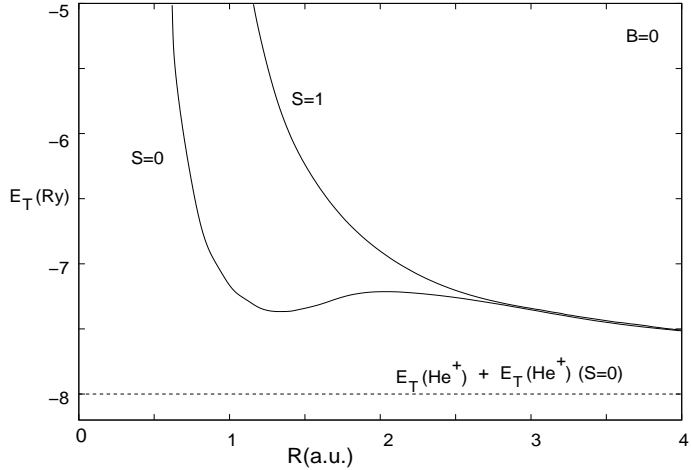}
    \caption{$B=0$: total energies of the ${}^1\Si_g\,(S = 0$, ground state)
    and ${}^3\Si_u\,(S = 1)$ states vs. the internuclear distance $R$.
    ${}^1\Si_g\,(S = 0)$ decays to two $He^+$ ions, their total energies
    are shown by dashed line.}
\end{center}
\end{figure}

%%%%%%%%%%%%%  FIGURE:2  %%%%
%%%%%%%%%%%%%%
\begin{figure}[tb]
\begin{center}
   \includegraphics*[width=3.in, height=4.in, angle=-90]{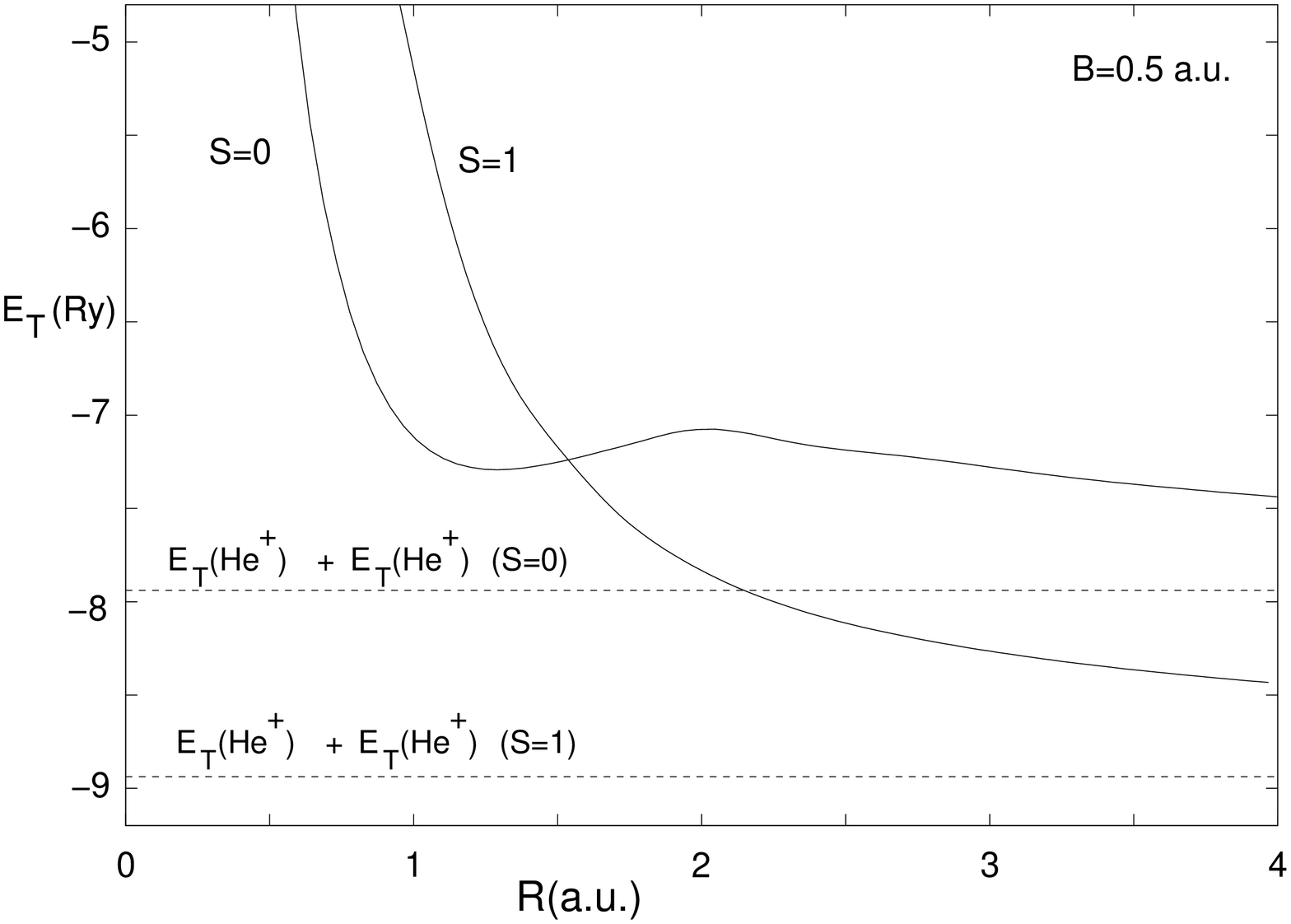}
    \caption{$B=0.5$: total energies of the ${}^1\Si_g\,(S = 0$, ground state)
    and ${}^3\Si_u\,(S = 1)$ states vs. the internuclear distance $R$.
    ${}^1\Si_g\,(S = 0)$ decays to two $He^+$ ions, their total
    energies for different total electronic spins are shown by dashed lines.}
\end{center}
\end{figure}

%%%%%%%%%%%%%  FIGURE:3  %%%%
%%%%%%%%%%%%%%
\begin{figure}[tb]
\begin{center}
   \includegraphics*[width=3.6in, height=3.in, angle=-90]{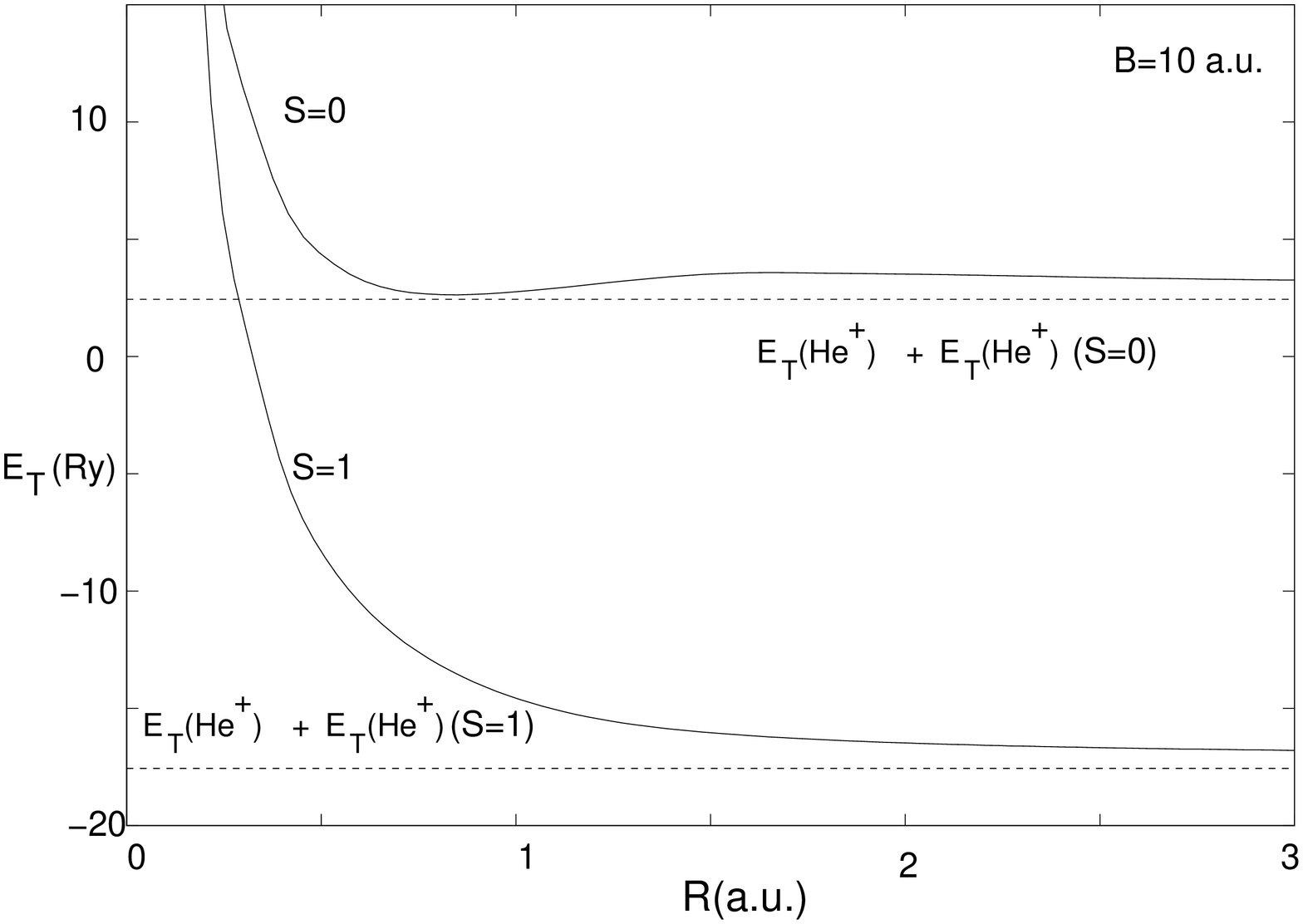}
    \caption{$B=10$: total energies of the ${}^1\Si_g\,(S = 0$)
    and ${}^3\Si_u\,(S = 1$\,, "ground state") states vs. the internuclear distance $R$.
    ${}^1\Si_g\,(S = 0)$ decays to two $He^+$ ions, their total
    energies for different total electronic spins are shown by dashed lines.}
\end{center}
\end{figure}

The performed variational calculations do not indicate to any
minimum in the total energy curve $E_T(R)$ of the $(\al \al e e)$
system at finite internuclear distances for magnetic fields ranging
$B=0 - 4.414 \times 10^{13}\,$\,G. Thus, the state ${}^3\Si_u$ is a
repulsive state for all the range of studied magnetic fields, $
0~\le~B~\le~4.414 \times 10^{13}\,$ G. One can state that this
system "exists" in a form of two Helium ions $He^+$ situated at
infinitely large distance from each other.

However, it can be seen that at $~B~ \approx ~0.85\,$ a.u. the
crossing occurs of the total energy curve at the equilibrium
internuclear distance for the  ${}^1\Si_g$ state and the (repulsive)
total energy curve of the ${}^3\Si_u$ state. It means that the
$He_2^{2+}$ molecular ion exists in the ${}^1\Si_g$ as a metastable
state for $ 0~\le~B~\le~0.85$\ a.u. However, for
$0~\ge~B~\le~0.85\,$a.u. the $He_2^{2+}$ ion ceases to exist as a
compact system characterized by a finite internuclear distance. But
it does exist in a form of two separate Helium ions $He^+$ with each
of them in $1s$ state with parallel electron spins. As an
illustration we present on Fig.1-3 the plots of total energies of
the ${}^1\Si_g$ and ${}^3\Si_u$ states for magnetic fields 0,
0.5\,a.u. and 10\,a.u.

As a next step the state $^3\Pi_u$ ($S = 1$, $\si_N =1, m=-1$) of
the $(\al \al e e)$ system in parallel configuration is studied in
the domain of magnetic fields $2 \times 10^9\,G \le B \le 4.414
\times 10^{13}\,$~G. For this state the variational trial function
(\ref{ansatz}) with $\si_e =-1$, $\si_N =1$ and $m=-1$ depends on 8
variational parameters. Obtained results (see Table~\ref{table2})
indicate to the existence of a clear minimum in the total energy
curve $E_T(R)$ for magnetic fields $B \ge 10$ \,a.u. With an
increase of the magnetic field strength the total energy at
equilibrium position decreases, the system becomes more bound (in
this case double ionization energy is equal to $E_I=-E_T$ and it
grows) and more compact (the internuclear equilibrium distance
decreases). Also, for magnetic fields $B \ge 10$ \,a.u. there exists
at least one longitudinal vibrational state (see
Table~\ref{table2}). Table~\ref{table2} also shows the total
energies of the dissociation channels $(\al \al e e) \to
He^+(1s_0)+He^+(2p_{-1})$, $(\al \al e e)  \to He_2^{3+}(1\si_g)+e$
and $(\al \al e e)  \to He(_3(-1)^{-1}) + \al$ for different
magnetic fields.

In Table~\ref{table2} the total energy of two Helium ions
$E_T(2)=2E_T(He^+(1s_0))$ is shown in the case of parallel spins of
electrons. This energy also corresponds to the lowest total energy
of the ${}^3\Si_u$ state (see above). The lowest total energies of
the ${}^3\Pi_u$ and $^3\Si_u$ states cross at $B \approx 1100
$\,a.u. At larger magnetic fields the lowest total energy of the
system $(\al \al e e)$ in the $^3\Pi_u$ state is lower than the
lowest total energy of the ${}^3\Si_u$ state (which correspond to
the separate ions $He^+$. Hence, the system $(\al \al e e)$
corresponds to the molecular ion $He_2^{2+}$ at $B \gtrsim 1100
$\,a.u. One can draw a conclusion that the molecular ion $He_2^{2+}$
{\it exists} at $B \gtrsim 1100 $\,a.u. It is a stable system where
its ground state is given by the $^3\Pi_u$ state. At smaller
magnetic fields 0.85\, a.u. $\lesssim \, B\, \lesssim 1100$\,a.u.
the lowest total energy of the system $(\al \al e e)$ corresponds to
the state $^3\Si_u$ at the infinite equilibrium distance. It means
that the system $(\al \al e e)$ does not form a bound state, it does
"exist" in a form of two separate Helium ions $He^+$.

If the system $(\al \al e e)$ can be kept externally in the
${}^3\Pi_u$ state, one can check a stability of the system towards
dissociation channels. A comparison of the total energy of $(\al \al
e e)$ in the ${}^3\Pi_u$ state and the total energy of the
(dominant) channel of decay $(\al \al e e)  \to
He^+(1s_0)+He^+(2p_{-1})$  indicates that the  $(\al \al e e)$
system in the ${}^3\Pi_u$ state becomes stable with respect to the
possible channels of decay for $B \gtrsim 100 $\,a.u. The
dissociation energy increases monotonously with a magnetic field
growth.

\squeezetable
\begin{table}
  \centering
\caption{\label{table2}
  The state ${}^3\Pi_u$: total $E_T$, lowest longitudinal vibrational
  $E_{vib}$ energies (in Ry) and equilibrium distance $R_{eq}$ (in a.u.)
  of the $He_2^{2+}$ ion as a function of the magnetic field.
  Magnetic fields for which $He_2^{2+}$ does not exist (see text) are marked by
  ${}^{(*)}$.
  Total energy of two Helium ions $He^+(1s_0) + He^+(2p_{-1})$
  with electron spins antiparallel to magnetic field (from \cite{Turbiner:2006})
  is denoted as $E_T(1)$, also the total energy of two Helium ions both in
  $1s$-state, $He^+(1s_0)$ each  of them with electron spin antiparallel
  to magnetic field from \cite{Turbiner:2006} is denoted as $E_T(2)$ for comparison as
  well as total energy of the system $He_2^{3+}(1\si_g)+e$ from
  \cite{Turbiner:2006}.
  Total energy $E_T(He(*))$ for $He$ atom in ${}_3(-1)^{-1})$-state
  was taken from \cite{schmelcherHe}. }
\begin{tabular}{c|c|c|c|c|c|c|c}
\hline\hline
 B(a.u.) & $E_T$  & $R_{eq}$ & $E_{vib}$ & $E_T(1)$  &
 $E_T(He_2^{3+}(1\si_g)+e)$ &$E_T(He(*))$ & $E_T(2)$  \\
\tableline
$10^{(*)}$      & -13.7694 & 1.100 & 0.03 & -14.018 &         & -11.67895  & -17.560 \\
%       &         &       &      &         &         &            &          \\
$100^{(*)}$     & -33.983  & 0.463 & 0.12 & -31.843 & -16.516 & -26.210    & -38.217  \\
%       &         &       &      &         &         &            &          \\
$1000^{(*)}$    & -80.492  & 0.212 & 0.40 & -69.150 & -39.268 &            & -80.801 \\
%       &         &       &      &         &         &            &          \\
10000   & -174.51 & 0.106 & 1.16 & -137.26 & -86.23  &            & -156.86 \\
%       &         &       &      &         &         &            &          \\
$B=4.414\times 10^{13}$\,G & -212.14 & 0.0902 &1.69  & -162.84 & -105.121 & & -185.06 \\
\tableline\hline
\end{tabular}
\\
\end{table}

To complete a study of the existence of the $He_2^{2+}$ ion we
carried out calculations of the energy for the state ${}^3\De_g$ ($S
= 1$, $\sigma_N =1 ,m=-2$) of the system $(\al \al e e)$ in parallel
configuration in the domain of magnetic fields 1\, a.u.$\,\le\, B\,
\le\, 4.414 \times 10^{13}\,$G.  For this state the variational
trial function (\ref{ansatz}) is used with $\si_e =-1$, $\si_N =1$
and $m=-2$, it depends on 8 variational parameters.

There exists a clear minimum in the total energy surface $E_T(R)$ of
$(\al \al e e)$ for magnetic fields $B \ge 10$ \,a.u. It is found
that with an increase of the magnetic field strength the total
energies decrease, the system becomes more bound (double ionization
energies increase, $E_I=-E_T$, for triplet states) and more compact
(the internuclear equilibrium distance decreases) (see Table
\ref{table3}). At $B \approx 7000$\,a.u. the crossing between the
total energy curves of the ${}^3\Si_u$  and ${}^3\Delta_g$ state
occurs. The $^3\Delta_g$ state becomes the lowest-lying excited
state for $B \gtrsim 7000 \,$a.u. until the Schwinger limit.

\begin{table}
  \centering
\caption{\label{table3}
       The state  $^3\Delta_g$: total energy $E_T$
       in Ry and equilibrium distance $R_{eq}$ in a.u. of
       the $He_2^{2+}$ ion.
       Magnetic fields for which $He_2^{2+}$ does not exist (see text)
       are marked by ${}^{(*)}$.}
\begin{tabular}{c|c|c}
\hline\hline
B(a.u.)    & $E_T$     & $R_{eq}$   \\
\hline
 $10^{(*)}$         & -12.3118  &  1.278  \\
%          &           &         \\
$100^{(*)}$    & -30.607   &  0.517  \\
%          &           &         \\
$1000^{(*)}$ & -73.125   &  0.230  \\
%          &           &         \\
10000      & -160.64   &  0.113  \\
%          &           &         \\
$B=4.414\times 10^{13}$\,G &-195.87 & 0.099  \\
\tableline\hline
\end{tabular}
\\
\end{table}

\section{Conclusion}

The existence of the $He_2^{2+}$ molecular ion and its low-lying
excited states in a magnetic field ranging from zero to $4.414
\times 10^{13}\,$ G in parallel configuration are studied in the
Born-Oppenheimer approximation using the variational method. It is
found that the quantum numbers of the state of lowest total energy
depend on the magnetic field strength. The ground state evolves from
the spin-singlet ${}^1\Si_g$ state which is a metastable state for
weak magnetic fields $B \lesssim 0.85$\,a.u. to the unbound
(repulsive) spin-triplet ${}^3\Si_u$ for intermediate fields
$0.85\,\mbox{a.u.} \lesssim \ B \lesssim 1100$\,a.u. and eventually
to a strongly bound spin-triplet ${}^3\Pi_u$ state for strong
magnetic fields $B \gtrsim 1100$\,a.u. It manifests the existence of
the stable $He_2^{2+}$ molecular ion at $B \gtrsim 1100$\,a.u.

In the domain $1100 \,$a.u. $\lesssim B \lesssim 7000 \,$a.u. where
the state ${}^3\Pi_u$ becomes the ground state, the lowest-lying
excited state is the repulsive ${}^3\Si_u$ state. The next excited
state is the "bound" $^3\Delta_g$ state (having a clear minimum in
the total energy curve) lies above the ${}^3\Si_u$ state. One can
state that the $He_2^{2+}$ molecular ion does not have excited
states in a traditional sense in this domain of magnetic fields. At
larger magnetic fields $B \gtrsim 7000 \,$a.u. up to the Schwinger
limit the lowest-lying excited state is ${}^3\De_g$, while the next
excited state is the repulsive ${}^3\Si_u$ state.

\begin{acknowledgments}
 The authors are grateful to J. C. Lopez Vieyra for helpful discussions
 and interest to the present work.
 This work was supported in part by FENOMEC and
 PAPIIT grant {\bf IN121106} (Mexico).
\end{acknowledgments}

\end{document}